# THE ESRF TANGO CONTROL SYSTEM STATUS

J-M Chaize, A. Götz, W-D. Klotz, J. Meyer, M. Perez, E. Taurel and P. Verdier
ESRF, BP220, Grenoble, 38043, FRANCE

Abstract

TANGO[1] is an object oriented control system toolkit based on CORBA[2] presently under development at the ESRF. In this paper, the TANGO philosophy is briefly presented. All the existing tools developed around TANGO will also be presented. This includes a code generator, a WEB interface to TANGO objects, an administration tool and an interface to LabView. Finally, an example of a TANGO device server for OPC device is given.

## 1 INTRODUCTION

The task of building a control system in today's world has been heavily influenced by the ever-increasing choice of Commodity off the Shelf products. Many of the control problems (hardware and software) have been solved and can be bought ready to use off-the-shelf. However the products have to be integrated in order to form a control system. System integration is therefore one of the main tasks of a control system builder today. TANGO has been developed with system integration as one of its main design goals.

In TANGO system integration is achieved by wrapping. Wrapping means inserting a layer of software between the product to be integrated and the system in which it has to be integrated. The wrapper layers runs on the product platform and communicates with the control system via the network. The wrapper software needs to be multi-platform, network based and language independent. TANGO has chosen CORBA as its wrapper software.

## 2 THE CORBA MIDDLEWARE

CORBA is what the software industry called a *middleware*. This means that it is a layer that allows end-user application to communicate with other end-user application or with utilities hiding all the communication protocol. The CORBA definition uses the object approach to deal with the communication problem. Object interfaces are defined using a language called IDL (Interface Definition Language). An object interface defines all the kind of requests that the object supports coming from the external world. CORBA defines language mappings from IDL to the main programming languages. Various commercial and non-commercial implementations exist for CORBA for all the mainstream operating systems.

## 3 TANGO PHILOSOPHY

The TANGO philosophy could be summarized in the following points:

- Hide CORBA details from the end-user application and from the device access software programmer. This is achieved by providing programmers with a Device pattern for implementing new control classes and an API[3] for implementing physics applications.
- Define only one type of network object. This means only a single IDL file and only a single type of object to support as far as the communication layer is concerned. All controlled objects will inherit from this base class. This ensures all objects support the same basic interface and functionality. This uniqueness of object control interface allows writing of generic application which can be used whatever the controlled device is.
- Group controlled objects in processes called *device server*s. All device server processes have the same architecture based on a well-defined device pattern.
- Keep a high degree of flexibility by using a database for device specific information and description.
- Use only freely available software

## 4 THE TANGO OBJECT INTERFACE (IDL)

Only one IDL file is defined as there is only one interface to support. Actions are performed on devices by executing commands. Each command is defined by its name and has one input and one output parameter (which could be void). These parameters must be one of a list of 20 data types supported by TANGO. For commands without input parameter, results are returned from the hardware itself or from a data cache continuously filled by a polling mechanism. Commands are sent to the device via a *command_inout* operation defined in the IDL. The use of CORBA Any object allows the same device operation to transmit different data types. Commands can be executed synchronously or asynchronously. Asynchronous commands have to supply a Callback object to receive the answer. Devices also support a list of attributes which could be read or write. Like commands, attribute value can be read from the hardware or from the data cache. Device also supports describing calls like *command_list_query* or *get_attribute_config* which enables generic application to deal with any kind of devices.

---

[1] TANGO – TAco Next Generation Object
[2] CORBA – Common Object Request Broker Architecture
[3] API – Application Programmer Interface

## 5 DEVICE SERVER PROCESS

A device server is an operating system process with one or several user implementations of the TANGO device pattern. Following a predefined *main()* or *winmain()* structure, the device access software programmer merges all the device pattern implementation s/he want to run within the same process. Device name and number for each device pattern implementation is defined in the database and is retrieved during the process startup sequence. Multiple instances of the same device server process may run within a single TANGO system. Each instance has an instance name specified when the process is started. The couple process executable name/instance name uniquely defines a device server.

## 6 TANGO STATUS

Since the last ICALEPCS meeting in Italy, the following features have been added to TANGO:
- Full support for Windows. It is now possible to run a TANGO device server in a MS-DOS window, as a stand-alone application with its own windows using MFC or the Win32 library, or as a service
- Full support for TANGO attributes
- Polling threads in device server process that enables data caching for slow devices.
- The Java and C++ API's
- A device server code generator called POGO
- A TANGO administration tool called ASTOR
- A WEB generic application to ease device testing called JIVE
- A LabView interface

These new features are described in the following chapters.

## 7 DATA TYPES AND API

TANGO supports a fixed set of data type for transferring data using commands. All simple types and sequences of simple types are supported. In addition, TANGO supports two mixed types (sequence of strings and longs, and a sequence of strings and doubles). For attributes, only four types of data are supported. These types are short, long, double or strings (in zero, one or two dimension arrays)

TANGO clients can be programmed using only the CORBA API. Nevertheless, CORBA knows nothing about TANGO and programming at this level means that clients have to know all the details about CORBA programming. The TANGO philosophy is to hide these recipes in an API. The API is implemented as a set of C++ classes or as a Java package. This API also implements automatic reconnection between clients and server in case of server restart or front-end computer reboot.

## 8 POGO: A TANGO DEVICE PATTERN GENERATOR

In order to ease the work of device access software programmer, a TANGO device pattern code generator has been written. This generator is called *pogo*. It is a graphical tool built using. Once commands, attributes, properties and device states has been clearly defined, it is possible to use pogo to generate a complete framework of the device pattern implementation. Using a user-friendly graphical interface, the programmer can enter all the commands, attributes, properties and states. The tool will generate C++ or Java classes and an HTML framework for documentation. Obviously, the code specific to the device is still in the programmer hands! Pogo is also able to analyze a device pattern implementation that it has not generated or to analyze a device pattern that has been modified by the programmer if s/he follow some very basic rules. This allows the usage of Pogo from the beginning to the end when developing a device pattern implementation.

## 9 JIVE: A GENERIC WEB TANGO DEVICE INTERFACE

Jive is a basic tool used by each TANGO programmer or user. It is made of two different parts, which are:
- A generic device menu. Once a device has been selected and, using device configuration commands, Jive displays the list of command/attribute supported by the device. The user is able to execute any command or to read/write any attribute.
- A graphical interface to the TANGO database. Jive offers a graphical interface to all the database functionalities like defining new properties, updating properties value, attaching a new device to a device server process, browsing devices and properties, etc

The two parts of this tool are written using Java and a servlet/applet couple. The graphical possibilities offered by HTML are not well adapted to the needs of such a program. Therefore, an applet has been written. The drawback of applet is its loading time and all its security restriction. In this architecture, the applet communicates with a servlet running within an Apache WEB server. Communication with the TANGO device is done by the servlet. The applet communicates with the servlet using the traditional doPost/doGet HTTP[4] request. This has a timing penalty but access time is not a key point when testing a device, checking its functionalities or browsing the database

## 10 ASTOR: ADMINISTRATING A COMPLETE TANGO CONTROL SYSTEM

Once the control system has several device servers, an administration tool is needed. Within TANGO toolkit, this tool is named *Astor*. It allows an easy starting/stopping of device server processes even on remote hosts. A specific device server called Starter achieves this. The starter device supports commands to start, stop device server process running on the same host. For a correct usage of Astor, each host involved in the control system must have a Starter device server.

For a correct start up/shutdown of complex TANGO control system, a level can be associated to each device server process. These levels are stored in the database. Starting or stopping the control system using Astor will ensure the sequencing of action according to these levels.

## 11 LABVIEW INTERFACE

TANGO has been interfaced to the LabView G language. A dynamically loaded library (dll) on Windows and a shared library on Unix has been written to convert the LabView G programming types to TANGO types. G programs use the shared library VI[5] to call TANGO. LabView programs written for one platform also run on the other platforms. The LabView TANGO interface is presently supported on Windows, Linux and Solaris.

## 12 EXAMPLE OF AN OPC DEVICE SERVER

Most of PLC[6] suppliers propose an OPC[7] interface with a data server (SDS) running on Windows NT. For a full distributed control system, we need to access this SDS easily from anywhere on our network. The idea was to write (with POGO pattern generator) a TANGO device server able to read/write data in PLC through the SDS.

This TANGO device pattern implementation gets the PLC's I/O addresses as properties from TANGO database. It could also be used as a low level device pattern implementation for a higher level device server for a specific PLC usage and/or controlling many PLCs distributed around the accelerator.

## 13 CONCLUSION

Nowadays, several TANGO device servers are used in the day to day running of the ESRF control system. With the help of the backward compatibility provided by the TACO API's, it was possible to smoothly incorporate them in our running control system.

TANGO is actually available on 4 platforms presently Linux (Suse), Windows NT, Solaris and HP-UX. C++ and Java are supported for client and/or server. It uses CORBA release 2.3.

The TANGO IDL has been written since the beginning of the project. Some of the features foreseen in this definition are still missing. This includes the security aspect, the grouped calls and the event system. This is our future task.

Even if CORBA has a steep learning curve, it is easy to use for building simple types of network objects like our Device. Performance of CORBA is more than enough for an object oriented control system. The paradigm of device oriented access has again proved to be very powerful and adopted to control system problems.

TANGO offers significant improvements compared to TACO e.g. its support for modern protocols (IIOP) and language (Java and C++), immediate reconnection, openness to emerging WEB technologies.

---

[4] HTTP – Hyper Text Transfer Protocol
[5] VI – Virtual Instrument
[6] PLC – Programmable Logic Controller
[7] OPC – OLE Process Control